\documentclass[12pt,onecolumn]{article}

\usepackage{graphicx}% Include figure files
\usepackage[utf8]{inputenc}
\usepackage{amsmath}
\usepackage{amsfonts}
\usepackage{amssymb}
\usepackage{graphicx}
\usepackage{bm}% bold math
\begin{document}
\title{Derivation of Van der Waal's equation of state in microcanonical ensemble formulation }
%\author{M. Ponmurugan}
\author{Aravind P. Babu, Kiran S. Kumar and M. Ponmurugan*  \\
Department of Physics, School of Basic and Applied Sciences, \\
Central University of Tamilnadu, Thiruvarur 610 005, Tamilnadu, India.} 
 %\affiliation{Department of Physics, School of Basic and Applied Sciences, Central University of Tamilnadu,
%Thiruvarur - 610 101, Tamilnadu, India}
%\date{\today}

%\begin{abstract}
%\end{abstract}

%\pacs{05.70.Ln,05.20.-y,05.70.-a}
%\pacs{
%05.20.-y - Classical statistical mechanics, 
%05.70.-a - Thermodynamics. 
%}
%\keywords{Van der Waal's gas, equation of state, microcanonical ensemble}
\maketitle

%\begin{abstract}
\section*{Abstract}
The Van der Waal's equation of state for a (slightly) non-ideal
classical gas is usually derived in the context of classical
statistical mechanics by using the canonical ensemble. We use the 
hard sphere potential with no short range interaction and derive Van 
der Waal's equation of state in microcanonical ensemble formulation. 
%\end{abstract}

\section{Introduction}
The ensemble theory of equilibrium statistical mechanics connects the macroscopic relations of thermodynamic systems and its microscopic constituents \cite{callen,pathria,huang}. An ensemble of a given thermodynamic system is a set of distinct microstates with appropriately assigned probability  for a fixed macrostate \cite{kpnmc}. In a macrostate of fixed energy $E$, volume $V$ and the number of particle $N$ and all its microstates are equiprobable then such an ensemble is called as microcanonical ensemble \cite{kpnst}. 

The most important concept in statistical mechanics theory is the equivalence of ensemble, that is, the ensemble theory 
should provide the same macroscopic relations in the thermodynamic limit of infinite size irrespective of the 
chosen ensemble. This concept of ensemble equivalence has been verified easily by incorporating 
different ensemble approach to simple system known as ideal gas. Whatever the ensemble one may choose, we can finally 
obtain the ideal gas equation of state as $PV=nRT$ \cite{pathria,kpnst}, where $P$
is the pressure, $V$ is the volume, $T$ is the temperature of the system,
$n$ is the number of moles and $R$ is the universal gas constant.

An ideal gas is the simplest thermodynamic  system in which there is no intermolecular interactions.  
By considering the intermolecular interactions, Van der Waals proposed the 
equation of state for a real gas which is given by \cite{sear}
\begin{eqnarray}
\left( P+\frac{n^{2}a}{V^2} \right) \left( V-nb \right) =nRT,
\label{van}
\end{eqnarray}
where $a$ and $b$ are the Van der Waal's constants.  
The  derivation of Van der Waal's equation of state has been found in most of the  
Statistical Physics books  which are in canonical ensemble formulation \cite{reif}.
To our knowledge, there is no such study available in the literature for any other ensemble formulation,
in particular, microcanonical ensemble formulation.
In this paper, we use the hard sphere potential approximation and derive 
the Van der Waal's equation of state in microcanonical ensemble formulation.

\section{Hamiltonian for  Van der Waal's gas}
Consider a monoatomic non ideal gas of $N$ particles having identical mass $m$ in a container of fixed volume $V$ at temperature $T$. In
order to treat the problem in classical statistical mechanics, we assume that the temperature is taken to be sufficiently high and the  density  $ \rho = \frac{N}{V}$ is sufficiently low. The total energy (Hamiltonian) of the system is $H=K+U$,
where $K=\frac{1}{2 m} \sum_{l=1}^{3N}p_{l}^{2}$ is the kinetic energy of the gas of $3N$ degrees of freedom moving with momentum $p_l$. $U$ is the total potential energy due to the interaction that exists between the molecules.  
In semi-classical approximation, we consider the molecules to be hard spheres of radius $r_0$, so that the distance between the two molecules can come close to $r_0$ \cite{pathria}. The interaction between a pair of molecules $i$ and $j$ separated by the intermolecular distance $r$ is given by the hard sphere potential with no short range interaction as \cite{reif,allen}
\begin{eqnarray}
 u(r) &=& -u_0 \bigg( \frac{r_0}{r} \bigg)^6 \ \mbox{for} \  r \ge r_0,
\label{hardpot}
\end{eqnarray}
where $u_0$ is the depth of the potential.
The total potential energy is given by the sum of interactions between the pair of all molecules as 
\begin{equation}
U = \frac{1}{2} \sum_{i=1}^{N} u_i,
\end{equation}
where $u_i$ is the interaction energy of $i^{th}$ molecule with all other molecules in the specified region given in spherical polar coordinates as 
\begin{equation}
u_{i}=\int_{r_0}^{\infty} \int_0^{2 \pi} \int_0^{\pi}  \rho u(r) \ r^{2} sin\theta \ dr d\theta d\phi  
\end{equation}
\begin{equation}
u_{i}=-4 \pi \frac{Nu_{0}}{V} \int_{r_{0}}^{\infty} \frac{r_{0}^6}{r^{4}} dr = - \frac{4 \pi N u_{0} r_{0}^3 }{3 V}.
\end{equation}
Here, we assume that the density $(\rho =\frac{N}{V})$ of the gas to be uniform throughout the volume. Thus,
the total potential energy is given by 
\begin{equation}
U=\frac{1}{2}N u_{i} =\frac{-a' N^{2}}{V},
\end{equation}
where $a'=\frac{2 \pi u_0 r_0^3}{3}$. Therefore the total Hamiltonian is given by 
\begin{equation}
H= \sum_{l=1}^{3N} \frac{p_{l}^{2}}{2m} - \frac{a^{'}N^{2}}{V} \equiv E.
\label{vanHE}
\end{equation}

\section{Microcanonical entropy and equation of state for Van der Waal's gas}

Because of the hard sphere approximation, there should be a correction in the total volume of the Van der Waal's gas.  
By considering the molecules as a hard sphere of diameter $2r_0$, the center of each molecules excluded 
by other molecule by a volume which is equivalent to the volume of sphere of radius $2r_0$ is known as the excluded volume.
The excluded volume for the two molecules of radius $r_0$ is $\frac{4}{3} \pi (2r_0)^3$. Since the system contains N molecules, 
the excluded volume $v$ for $N$ molecule can be obtained as $v=Nb'$ where $b'=\frac{2 \pi (2r_0)^{3}}{3}$.
The corrected volume which is available for the gas molecules in the container is given by
\begin{equation}
 V'=V-Nb'.
\end{equation}

Consider a small volume in phase space, the total number of microstates available for the system in 
microcanonical ensemble of fixed $E,V$ and $N$ of Eq.(\ref{vanHE}) is given by \cite{pathria,huang,sear} 
\begin{equation}
\Omega(E,V,N)=\frac{1}{N ! h^{3N}}\frac{\partial \omega}{\partial E},
\label{nomic}
\end{equation}
where $h$ is the Planck's constant and the volume integral \cite{kpnst,sear,greiner}
\begin{eqnarray}
 \omega(E,V,N)&=& \int \int_{H(q,p) \leqslant E} d^{3N}q\; d^{3N}p.
\label{volint0}
 \end{eqnarray} 
For hard sphere potential, the above integral can be written as
\begin{eqnarray}
 \omega(E,V,N)&=& ( V-Nb' )^N \int_{H(q,p) \leqslant E}  d^{3N}p,
\label{volint}
 \end{eqnarray} 
where $\int_{H(q,p) \leqslant E} d^{3N}q=(V')^N=(V-Nb')^N $ for Van der Waal's gas and 
Eq.(\ref{vanHE}) can be rearranged as
\begin{equation}
  \sum_{l=1}^{3N} p_l^2  = 2m \bigg( E+\frac{a'N^2}{V} \bigg)  
\end{equation}
The integral in Eq.(\ref{volint}) is just the volume of a $3N$ dimensional
sphere of radius $R=\sqrt{2m(E+\frac{a'N^2}{V})}$ which can be obtained as
(see appendix) \cite{pathria,huang}
\begin{eqnarray}
\omega(E,V,N)&=&\frac{\pi^{3N/2}}{\frac{3N}{2}\Gamma{(\frac{3N}{2})}}(V-Nb')^NR^{3N} \\
  &=&\frac{\pi^{3N/2}}{\frac{3N}{2}\Gamma{(\frac{3N}{2})}}(V-Nb')^N \Bigg[2m \bigg(E+\frac{a'N^2}{V}\bigg) \Bigg]^{3N/2}
\end{eqnarray}
Using the above equation we can calculate the number of microstates as in Eq.(\ref{nomic}) for Van der Waal's gas 
in micro canonical ensemble as 
\begin{equation}
\Omega(E,V,N)=\frac{1}{N ! h^{3N}}\frac{\pi^{3N/2}}{\Gamma{(\frac{3N}{2})}}(V-Nb')^N(2m)^{3N/2}\bigg(E+\frac{a'N^2}{V} \bigg)^{\frac{3N}{2}-1}.
\end{equation}
For large $N$, we can approximate
$\frac{3N}{2}-1 \simeq 3N/2$ and $\Gamma(\frac{3N}{2})=(\frac{3N}{2}-1)! \simeq \frac{3N}{2}!$
then
\begin{equation}
\Omega(E,V,N)=\frac{1}{N ! h^{3N}}\frac{\pi^{3N/2}}{\frac{3N}{2}!}(V-Nb')^N(2m)^{3N/2}\bigg(E+\frac{a'N^2}{V}\bigg)^{\frac{3N}{2}}.
\label{nomicf}
\end{equation}
The Boltzmann microcanonical entropy of the system is given by \cite{pathria,kpnst,sear},
\begin{equation}
S(E,V,N)=k_B \ \ell n \Omega(E,V,N).
\end{equation}
Using Eq.(\ref{nomicf}) and applying Stirling approximation $\ell n N! = N \ell n N -N $, we can obtain the microcanonical 
entropy of Van der Waal's gas is 
\begin{equation}
S(E,V,N) = k_B N \Bigg\{ \frac{5}{2}+ln\Bigg\{\bigg(\frac{V-Nb'}{N} \bigg) \Bigg[\frac{4 \pi m }{3Nh^2} \bigg(E+ \frac{a'N^2}{V} \bigg) \Bigg]^{\frac{3}{2}}\Bigg\} \Bigg\}.
\label{ent}
\end{equation}

 At constant $V$ and $N$, from the Maxwell's First thermodynamic relation $dE=TdS - pdV$, we get 
\cite{callen,kpnst}
\begin{equation}
\frac{\partial S}{\partial E} =\frac{1}{T}
\end{equation}
\begin{equation}
\frac{\partial S}{\partial V}=\frac{P}{T}
\end{equation}
Using Eq.(\ref{ent}), we obtain
\begin{equation}
\frac{1}{T} = \frac{3}{2} N k_B  \bigg(E+ \frac{a'N^2}{V} \bigg)^{-1}
\end{equation}
The above equation can be rearranged in terms of energy as
\begin{equation}
E=\frac{3}{2} N k_B T - \frac{a'N^2}{V}
\label{ave}
\end{equation}
Using Eq.(\ref{ent}) and Eq.(\ref{ave}), one can get
\begin{equation}
\frac{P}{T}=\frac{Nk_B}{(V-Nb')}-\frac{a'N^2}{V^2T}.
\end{equation}
\begin{equation}
\bigg(P+\frac{a'N^2}{V^2}\bigg)(V-Nb')=Nk_BT
\end{equation}
Substitute $N=nN_a$, $a=a'N_a^2$, $b=b'N_a$ and $k_BN_a=R$ in the above equation, 
we get
\begin{equation}
\bigg(P+\frac{n^2a}{V^2}\bigg)(V-nb)=nRT.
\end{equation}
where R is gas constant, $n$ is the number of moles 
and $N_a$ is the Avogadro's number.
Thus, we have obtained  Van der Waal's equation of state with Van der Waal's constants a and b.

\section{Conclusion}
We used the hard sphere potential and derived
Van der Waal's equation of state in microcanonical ensemble formulation. 
Even though the canonical ensemble is much easier to use in actual application
than the microcanonical ensemble, one should not ruled out the various studies of 
the same system in other ensemble formulation. In this context, our result also verified the equivalence of ensemble for Van der Waal's gas. 

\section*{Appendix}
The integral in Eq.(\ref{volint}) can be considered as the volume of  $3N$ dimensional spheres 
of radius $R=\sqrt{2m(E+\frac{a'N^2}{V})}$ which is given by \cite{kpnst,sear}
\begin{eqnarray}
  V_N(R)&=& \int_{\sum_{l=1}^{3N} p_l^2 \leqslant R^2}d^{3N}p   \\
	      &=&R^{3N}\int_{\sum_{l=1}^{3N} y_l^2 \leqslant 1}d^{3N}y  \\
				&=& R^{3N} C_{3N},
\end{eqnarray}
where $y_l=p_l/R$, $C_{3N}=\int_{\sum_{l=1}^{3N} y_l^2 \leqslant 1}d^{3N}y$ and 
\begin{equation}
  d^{3N}y= dV_N(R) = 3N C_{3N} R^{3N-1} dR
\label{dvN}
\end{equation}
Using the identity $\int_{- \infty}^{+ \infty} e^{-y^{2}} dy = \sqrt{\pi}$ and Eq.(\ref{dvN}), we consider the integral
 \begin{equation}
\int_{- \infty}^{+ \infty} . . . . \int_{- \infty}^{+ \infty} e^{-(y_1^{2}+y_2^{2}+.......+y_{3N}^{2})} d^{3N}y = \pi^{\frac{3N}{2}}
 \end{equation}
 and transform it in to polar coordinates as,
\begin{equation}
 3NC_{3N}\int_0^{\infty} R^{3N-1}e^{-R^2}dR=\pi^{\frac{3N}{2}}.
 \end{equation}
Substituting $ R^{2}=x$, we obtain
\begin{eqnarray}
C_{3N}=\frac{\pi^{3N/2}}{\frac{3N}{2}\Gamma{(\frac{3N}{2})}}
\end{eqnarray}
where the Gamma function 
\begin{equation} 
 \Gamma{\bigg(\frac{3N}{2} \bigg)}= \int_{0} ^{\infty} x^{\frac{3N}{2}-1}e^{-x}dx.
\end{equation}
Therefore, Eq.(\ref{volint}) becomes
\begin{eqnarray}
\omega(E,V,N)&=&\frac{\pi^{3N/2}}{\frac{3N}{2}\Gamma{(\frac{3N}{2})}}(V-Nb')^NR^{3N} \\
  &=&\frac{\pi^{3N/2}}{\frac{3N}{2}\Gamma{(\frac{3N}{2})}}(V-Nb')^N \Bigg[2m \bigg(E+\frac{a'N^2}{V}\bigg) \Bigg]^{3N/2}
\end{eqnarray}

\paragraph*{Acknowledgements}
I would like to thank N. Barani Balan for proof reading the manuscript.

\paragraph*{email correspondence: ponphy@cutn.ac.in}

\end{document}